\documentstyle[aps,pra,eqsecnum]{revtex}
\begin{document}
\bibliographystyle{unsrt}
\title {A relativistic superalgebra in a  generalized Schr\"odinger  picture}  
\date{\today}
\author{
R. A. Frick\thanks{Email: rf@thp.uni-koeln.de}\\{\it Institut f\"ur Theoretische Physik, Universit\"at K\"oln, D-50923 Cologne, Germany }  \\}

\maketitle
\begin{abstract}
We  consider  a relativistic superalgebra  in the  picture in which the  time and spatial derivative  cannot be presented in the operators of the particle.  The supersymmetry generators as well as  the Hamilton operators  for the massive relativistic particles  with spin zero and spin-1/2 are expressed in terms of the principal series of the  unitary representations of the Lorentz group.  We also consider  the massless case.  New Hamilton  operators are conctructed  for the  massless particles with spin zero and spin 1/2.
\end{abstract}
\section{Introduction}
The aim of the present paper is to construct a relativistic supersymmetry algebra in   the  generalized Schr\"odinger picture which has been  proposed  in \cite{Frick1}.  In this generalization the analogue of Schr\"odinger operators of the particle are independent of both time and space coordinates ${t},{\bf x}$. The  derivatives ${\partial}_{t}$ and $\nabla_{\bf x}$  cannot be presented in these operators. This picture is based on the principal series of the unitary representations of the Lorentz group. The  non unitary  representations  are not useful in the  generalized Schr\"odinger picture.  For a supersymmetric model in this approach it is necessary to develop a new mathematical formalism  in which the supersymmetry generators are expressed in terms  of the  principal series of the spacetime independent representations of the Lorentz group.

 The principal series  correspond to the eigenvalues $1+\alpha^2-{\lambda}^2,\,(0\leq\alpha<\infty,\quad\lambda=-s,...,s,{\quad}s = spin)$) of the first $C_1={\bf N}^2-{\bf J}^2$, (${\bf N},\,{\bf J}$ are boost and rotation generators) and the eigenvalues $\alpha\lambda$ of the second Casimir operator $C_2={\bf N}\cdot{\bf J}$ of the Lorentz algebra \cite{Joos,Lom,Barut}. For  a  particle with spin zero  as the eigenfunctions of  $C_1$ in the momentum representation (${\bf p}$ = momentum, $p_0=\sqrt{m^2+{\bf p}^2}$, m=mass) one can choose  the functions (we use here the notation of references \cite{Frick1})
\begin{equation}
\label{1.1}
\xi^{(0)}({\bf p},{\alpha},{\bf n})=\frac{1}{(2\pi)^{3/2}}[(pn)/m]^{-1+i\alpha},
\end{equation} 
where  $n$ is a  vector on the  light-cone $n^2_0-{\bf n}^{2}=0$. 
For a  particle with spin-1/2  the eigensolutions  of  both operators $C_1$ and $C_2$ may be written in the form 
\begin{equation}
\label{1.2}
\widetilde{\xi}^{(1/2)}({\bf p},{\alpha},{\bf n})=D^{(1/2)}({\bf p},{\bf n})\,D^{(1/2)}({\bf n})\,{\xi}^{(0)}({\bf p},{\alpha},{\bf n}),
\end{equation}
where
\begin{equation}
\label{1.3}
{D}^{(1/2)}({\bf p},{\bf n})=\frac{pn+m-i{\vec \sigma}\cdot({\bf p}{\times}{\bf n})}{\sqrt{2(p_0+m)(pn)}},\quad{D}^{\dag(1/2)}({\bf p},{\bf n})\,{D}^{(1/2)}({\bf p},{\bf n})=1,
\end{equation}
and the matrix $D^{(1/2)}({\bf n})$ contains the eigenfunctions of the operator ${\vec \sigma}\cdot{\bf n}$ ($D^{\dag(1/2)}({\bf n})D^{(1/2)}({\bf n})=1$). 

From the point of view of a  supersymmetric model the matrices  $D^{(1/2)}({\bf p},{\bf n})\,D^{(1/2)}({\bf n})$ in (\ref{1.2}) and   $D^{\dag(1/2)}({\bf n}){D}^{\dag(1/2)}({\bf p},{\bf n})$ in
\begin{equation}
\label{1.4}
D^{\dag(1/2)}({\bf n}){D}^{\dag(1/2)}({\bf p},{\bf n})\widetilde\xi^{(1/2)}({\bf p},{\alpha},{\bf n})=\xi^{(0)}({\bf p},{\alpha},{\bf n}),
\end{equation}
may be regarded as matrices which realize  supersymmetry transformations. In this paper we use these matrices  to construct a supersymmetry algebra in  terms of the group parameter ${\alpha}$ and the  vector on the  light-cone ${\bf n}$. The paper is set out as follows. First  we quote the necessary results from the  Poincar\'e algebra for the particles with  spin zero and spin-1/2  in the ${\alpha}\,{\bf n}$ representation. In the third section  in  this representation the supersymmetry generators are constructed.  In the fourth section the massless case is  considered. In this section the supersymmetry generators will be used for the construction of  the Hamilton  and momentum operators for the massless particles with spin zero and spin 1/2 in the ${\alpha}\,{\bf n}$ representation.
\section{The Poincar\'e algebra} 

The plane waves $\sim\exp[-ixp]$   in the states in the generalized Schr\"odinger picture   appear in different representations. There is no ${\bf x}$ representation. Consequently, the spatial derivative $-i\nabla_{\bf x}$  cannot be used to construct  the Hamilton and the momentum operators  of the particle.  For these operators in this approach one must use a spacetime independent representation. Here for the massive relativistic particles with spin zero and spin-1/2  we use  the following operators.
In \cite{Kad1} it was shown that  the functions $({\bf n}=(\sin{\theta}\cos{\varphi},\sin{\theta}\sin{\varphi},\cos{\theta}))$
\begin{equation}
\label{2.1}
\xi^{\ast(0)}({\bf p},{\alpha},{\bf n})=\frac{1}{(2\pi)^{3/2}}[(pn)/m]^{-1-i\alpha},
\end{equation}
are  the eigenfunctions of the differential-difference operators (${\bf L}({\bf n}):={\bf L}$)
\begin{equation}
\label{2.2}
H^{(0)}=m\lbrack{\cosh(i\frac{\partial}{{\partial}{\alpha}})+\frac{i}{{\alpha}}\sinh(i\frac{\partial}{{\partial}{\alpha}}) +\frac{{\bf L}^2}{2{\alpha}^2}\exp(i\frac{\partial}{{\partial}{\alpha}})}\rbrack,
\end{equation}  
\begin{equation}
\label{2.3}
{\bf P}^{(0)}={\bf n}\lbrack{H^{(0)}-m\exp(i\frac{\partial}{{\partial}{\alpha}})}\rbrack-m\frac{{\bf n}{\times}{\bf L}}{\alpha}\exp(i\frac{\partial}{{\partial}{\alpha}}).
\end{equation}
These operators  satisfy the  conditions
\begin{equation}
\label{2.4}
  H^{(0)}\,\xi^{\ast(0)}({\bf p},{\alpha},{\bf n})=p_{0}\,\xi^{\ast(0)}({\bf p},{\alpha},{\bf n}),\quad{\bf P}^{(0)}\,{\xi}^{\ast(0)}({\bf p},{\alpha},{\bf n})={\bf p}\,{\xi}^{\ast(0)}({\bf p},{\alpha},{\bf n}).    
\end{equation}
                            
In \cite{Frick2}, in order to construct  the Hamilton  $H^{(1/2)}$ and the momentum operators  ${\bf P}^{(1/2)}$  for a  relativistic particle with spin 1/2 in the ${\alpha}\,{\bf n}$ representation the functions (${\xi}^{(1/2)}({\bf p},{\alpha},{\bf n})$ are the eigenfunctions of $C_1$)
\begin{equation}
\label{2.5}
\xi^{\dag(1/2)}({\bf p},{\alpha},{\bf n})=D^{\dag(1/2)}({\bf p},{\bf n})\xi^{\ast(0)}({\bf p},{\alpha},{\bf n}),
\end{equation}
and  the conditions
\begin{equation}
\label{2.6}
H^{(1/2)}\,\xi^{\dag(1/2)}({\bf p},{\alpha},{\bf n})=p_{0}\,\xi^{\dag(1/2)}({\bf p},{\alpha},{\bf n}),\quad{\bf P}^{(1/2)}\,{\xi}^{\dag(1/2)}({\bf p},{\alpha},{\bf n})={\bf p}\,{\xi}^{\dag(1/2)}({\bf p},{\alpha},{\bf n}),   
\end{equation}
were used.
The operators  $H^{(1/2)}$ and ${\bf P}^{(1/2)}$ have the form (${\bf J}^{(1/2)}={\bf L}+{\vec \sigma}/2$)
\begin{eqnarray}
\label{2.7}
H^{(1/2)}=\frac{m}{2}[(\frac{\alpha(\alpha+\imath)+({\bf J^{(1/2)}})^2}{({\alpha}^2+\frac{1}{4})})\exp(i\frac{\partial}{{\partial}{\alpha}})+\frac{\alpha-\frac{3i}{2}}{(\alpha-\frac{i}{2})}\exp(-i\frac{\partial}{{\partial}{\alpha}})-\frac{{\vec \sigma}\cdot{\bf L}+1}{(\alpha^2+\frac{1}{4})}],
\end{eqnarray}
\begin{eqnarray}
\label{2.8}
{\bf P}^{(1/2)}&=&{\bf n}[H^{(1/2)}-m\exp(i\frac{\partial}{{\partial}{\alpha}})]+m[\frac{{\bf n}{\times}{\vec \sigma}}{2(\alpha+i/2)}\nonumber\\&&-\frac{2\alpha({\bf n}{\times}{\bf L})+(\alpha-i/2){\bf n}{\times}{\vec \sigma}+({\bf n}{\vec \sigma}){\bf L}}{2(\alpha^2+1/4)}\exp(i\frac{\partial}{{\partial}{\alpha}})].
\end{eqnarray}

If in addition to  $H^{(s)}$ and  ${\bf P}^{(s)}$ (s=0,1/2) we use the operators of the Lorentz algebra
\begin{equation} 
\label{2.9}
{\bf J}^{(0)}:={\bf L},\quad{\bf N}^{(0)}={\alpha}{\bf n}+({\bf n}\times{\bf L}-{\bf L}\times{\bf n})/2,
\end{equation}
\begin{equation} 
\label{2.10}
{\bf J}^{(1/2)}={\bf L}+\frac{\vec \sigma}{2},\quad{\bf N}^{(1/2)}={\alpha}{\bf n}+({\bf n}\times{\bf J}^{(1/2)}-{\bf J}^{(1/2)}\times{\bf n})/2,
\end{equation}
 then we have the Poincare algebra  in the spacetime independent ${\alpha}$\,${\bf n}$ representation
\begin{eqnarray}
\label{2.11}
\lbrack{N^{(s)}_i},{P^{(s)}_j}\rbrack=\imath\delta_{ij}{H^{(s)}},\quad\lbrack{P^{(s)}_i},{H^{(s)}}\rbrack=0,\quad\lbrack{H^{(s)}},{N^{(s)}_i}\rbrack=-\imath{P^{(s)}_i},
\end{eqnarray}
\begin{eqnarray}
\label{2.12}
\lbrack{P^{(s)}_i},{P^{(s)}_j}\rbrack=0,\quad\lbrack{J^{(s)}_i},{H^{(s)}}\rbrack=0,\quad\lbrack{P^{(s)}_i},{J^{(s)}_j}\rbrack=\imath\epsilon_{ijk}{P^{(s)}_k},
\end{eqnarray}
\begin{eqnarray}
\label{2.13}
\lbrack{J^{(s)}_i},{J^{(s)}_j}\rbrack=\imath\epsilon_{ijk}{J^{(s)}_k},\quad
\lbrack{N^{(s)}_i},{N^{(s)}_j}\rbrack=-\imath\epsilon_{ijk}{J^{(s)}_k},\quad\lbrack{N^{(s)}_i},{J^{(s)}_j}\rbrack=\imath\epsilon_{ijk}{N^{(s)}_k}.
\end{eqnarray} 
For this reason  the operators $H^{(0)}$,\, ${\bf P}^{(0)}$ and $H^{(1/2)}$,\, ${\bf P}^{(1/2)}$  in the  generalized Schr\"odinger picture can be indentified with the Hamilton  and momentum operators for the  massive relativistic particles with  spin zero and spin-1/2, respectively.

In order to define the Hamilton  and the momentum operators for the spin-1/2 particle one can also use the functions
\begin{equation}
\label{2.14}
\widetilde{\xi}^{\dag(1/2)}({\bf p},{\alpha},{\bf n})=D^{\dag(1/2)}({\bf n})\xi^{\dag(1/2)}({\bf p},{\alpha},{\bf n}).
\end{equation}
On the basis of these functions 
\begin{equation}
\label{2.15}
\widetilde{H}^{(1/2)}\,\widetilde{\xi}^{\dag(1/2)}({\bf p},{\alpha},{\bf n})=p_{0}\,\widetilde\xi^{\dag(1/2)}({\bf p},{\alpha},{\bf n}),\quad\widetilde{\bf P}^{(1/2)}\,\widetilde{\xi}^{\dag(1/2)}({\bf p},{\alpha},{\bf n})={\bf p}\,\widetilde{\xi}^{\dag(1/2)}({\bf p},{\alpha},{\bf n}),    
\end{equation}
where
\begin{equation}
\label{2.16}
\widetilde{H}^{(1/2)}=D^{\dag(1/2)}({\bf n}){H}^{(1/2)}D^{(1/2)}({\bf n}),\quad\widetilde{\bf P}^{(1/2)}=D^{\dag(1/2)}({\bf n}){\bf P}^{(1/2)}D^{(1/2)}({\bf n}).    
\end{equation}
In the Poincar\'e algebra in this case instead of ${\bf J}^{(1/2)},\,{\bf N}^{(1/2)}$ we have
\begin{equation}
\label{2.17}
\widetilde{\bf J}^{(1/2)}=D^{\dag(1/2)}({\bf n}){\bf J}^{(1/2)}D^{(1/2)}({\bf n}),\quad\widetilde{\bf N}^{(1/2)}=D^{\dag(1/2)}({\bf n}){\bf N}^{(1/2)}D^{(1/2)}({\bf n}).    
\end{equation}
Bellow we use  $H^{(0)},\,{\bf P}^{(0)},\,{\bf J}^{(0)}$,\,${\bf N}^{(0)}$ and  $\widetilde{H}^{(1/2)},\,\widetilde{\bf P}^{(1/2)},\,\widetilde{\bf J}^{(1/2)},\,\widetilde{\bf N}^{(1/2)}$ to construct the supersymmetry generators.
\section{supersymmetry generators}

In \cite{Frick2} the Hamilton operator $H^{(1/2)}$ was constructed with help of the operator
\begin{equation}
\label{3.1}
B=\sqrt{m}\left[2\cosh(\frac{i}{2}\frac{\partial}{\partial\alpha})-\frac{i{\vec \sigma}\cdot{\bf L}}{(\alpha-i/2)}\exp(\frac{i}{2}\frac{\partial}{\partial\alpha})\right].
\end{equation}
This operator  was obtained  by means of replacing in the matrix $({\sqrt{2(p_0+m)}})D^{\dag(1/2)}({\bf p},{\bf n})$ the  quantity  ${p_0}$ by $H^{(0)}$ and the  quantities ${\bf p}$  by  ${\bf P}^{(0)}$
\begin{equation}
\label{3.2}
\xi^{\dag(1/2)}({\bf p},{\alpha},{\bf n})={B}\frac{\xi^{\ast(0)}({\bf p},{\alpha},{\bf n})}{{\sqrt{2(p_0+m)}}}.
\end{equation}
Another operator for which
\begin{equation}
\label{3.3}
KB=2(H^{(0)}+m),\quad{BK}=2({H^{(1/2)}}+m),
\end{equation}
has the form
\begin{equation}
\label{3.4}
K=\sqrt{m}\left[2\cosh(\frac{i}{2}\frac{\partial}{\partial\alpha})+\frac{2i}{\alpha}\sinh(\frac{i}{2}\frac{\partial}{\partial\alpha})+\frac{i{\vec \sigma}\cdot{\bf L}}{\alpha}\exp(\frac{i}{2}\frac{\partial}{\partial\alpha})\right].
\end{equation}
Introducing the  operators
\begin{equation}
\label{3.5}
\widetilde{B}=D^{\dag(1/2)}({\bf n})B,\quad\widetilde{K}=KD^{(1/2)}({\bf n}),
\end{equation}
we obtain
\begin{equation}
\label{3.6}
\widetilde{K}\widetilde{B}=2(H^{(0)}+m),\quad\widetilde{B}\widetilde{K}=2({\widetilde{H}^{(1/2)}}+m).
\end{equation}
Considering the eigenfunctions of $H^{(0)}$ and $\widetilde{H}^{(1/2)}$  as superpartner one can define  $\widetilde{B}$ and $\widetilde{K}$ as operators which realize the following supersymmetry transformations
 \begin{equation}
\label{3.7}
\widetilde{B}\,{\xi}^{\ast(0)}({\bf p},{\alpha},{\bf n})= {\sqrt{2(p_0+m)}}\widetilde{\xi}^{\dag(1/2)}({\bf p},{\alpha},{\bf n}),
\end{equation}
\begin{equation}
\label{3.8}
\widetilde{K}\widetilde{\xi}^{\dag(1/2)}({\bf p},{\alpha},{\bf n})={\sqrt{2(p_0+m)}}\xi^{\ast(0)}({\bf p},{\alpha},{\bf n}).
\end{equation}
Using the anticommuting operators
\begin{displaymath}
{Q_1}=\left( \begin{array}
{ccc}0 & \widetilde{K}\\\widetilde{B}& 0  
\end{array} \right),\quad{Q_2}=\left( \begin{array}
{ccc}0 &\imath\widetilde{K}\\-\imath\widetilde{B}& 0  
\end{array} \right),
\end{displaymath}
we have the relations
\begin{equation}
\label{3.9}
{Q_1}^2={Q_2}^2=2(H+m),
\end{equation}
\begin{equation}
\label{3.10}
\lbrack{H,Q_1}\rbrack=0,\quad\lbrack{H,Q_2}\rbrack=0,
\end{equation}
where 
\begin{displaymath}
H:=\left( \begin{array}
{ccc}H^{(0)} & 0\\ 0 & \widetilde{H}^{(1/2)}
\end{array} \right).
\end{displaymath}
With help of ${Q_1},\,{Q_2}$ and 
\begin{displaymath}
{\bf J}:=\left( \begin{array}
{ccc}{\bf J}^{(0)} & 0\\ 0 & \widetilde{\bf J}^{(1/2)}
\end{array} \right),\quad{\bf N}:=\left( \begin{array}
{ccc}{\bf N}^{(0)} & 0\\ 0 & \widetilde{\bf N}^{(1/2)}
\end{array} \right),
\end{displaymath}
one can construct other supersymmetry generators. The generators
\begin{eqnarray}
\label{3.11}
{Q_{1i}}:=\lbrack{Q_1,J_{i}}\rbrack,\quad{Q_{2i}}:=\lbrack{Q_2,J_{i}}\rbrack,
\end{eqnarray}
 may be expressed in the form
\begin{displaymath}
{Q_{1i}}=\left( \begin{array}
{ccc}0 &{\sigma}_i\widetilde{K}/2\\-\widetilde{B}{{\sigma}_i}/2 & 0 
\end{array} \right),\quad{Q_{2i}}=\left( \begin{array}
{ccc}0 &\imath{\sigma}_i\widetilde{K}/2\\\imath\widetilde{B}{{\sigma}_i}/2 & 0 
\end{array} \right),
\end{displaymath}
from which we can find that ($r=1,2$)
\begin{equation}
\label{3.12}
\{Q_{ri},Q_{rk}\}=-(H+m){\delta}_{ik},
\end{equation}
\begin{equation}
\label{3.13}
\lbrack{Q_{r1},J_{1}}\rbrack=\frac{1}{4}Q_{r},\quad\lbrack{H,Q_{ri}}\rbrack=0,
\end{equation}
and  for $i\not=j\not=k$
\begin{equation}
\label{3.14}
\lbrack{Q_{ri},J_{j}}\rbrack=\frac{\imath}{2}\epsilon_{ijk}Q_{rk}.
\end{equation}
For the commutators $\lbrack{Q_1,N_{i}}\rbrack$ and $\lbrack{Q_2,N_{i}}\rbrack$ \begin{eqnarray}
\label{3.15}
\lbrack{Q_1,N_{i}}\rbrack:={G_{1i}},\quad\lbrack{Q_2,N_{i}}\rbrack:={G_{2i}},
\end{eqnarray}
we obtain the relations
\begin{equation}
\label{3.16}
\{G_{ri},G_{rk}\}=-(H-m){\delta}_{ik},
\end{equation}
\begin{displaymath}
\{Q_{r},G_{ri}\}=-2\imath\left( \begin{array}
{ccc}P^{(0)}_i & 0\\ 0 &\widetilde{P}^{(1/2)}_i
\end{array} \right):=-2i{P_i},
\end{displaymath}
\begin{equation}
\label{3.17}
\lbrack{H,G_{ri}}\rbrack=0,\quad\lbrack{{\bf P},Q_{ri}}\rbrack=0,\quad\lbrack{{\bf P},G_{ri}}\rbrack=0,\quad\lbrack{G_{r1},N_{1}}\rbrack=-\frac{1}{4}Q_{r},
\end{equation}
and ($i\not=j\not=k$)
\begin{equation}
\label{3.18}
\lbrack{G_{ri},N_{j}}\rbrack=-\frac{\imath}{2}\epsilon_{ijk}Q_{rk},\quad\lbrack{G_{ri},J_{j}}\rbrack=\frac{\imath}{2}\epsilon_{ijk}G_{rk},\quad\lbrack{Q_{ri},N_{j}}\rbrack=\frac{\imath}{2}\epsilon_{ijk}G_{rk}.
\end{equation}
We write down the explicit form of ${G_{1i}}$
\begin{displaymath}
{G_{1i}}=\left( \begin{array}
{ccc}0 &\sqrt{m}[({g_{1i}})^{+}_{12}+({g_{1i}})^{-}_{12}]D^{(1/2)}({\bf n})\\{\sqrt{m}}D^{\dag(1/2)}({\bf n})[({g_{1i}})^{+}_{21}+({g_{1i}})^{-}_{21}]& 0  
\end{array} \right),
\end{displaymath}
where
\begin{equation}
\label{3.19}
({g_{1i}})^{+}_{12}=\frac{[iN^{(0)}_i+(\vec \sigma\times{\bf N}^0)_i]}{2\alpha}{\exp(\frac{i}{2}\frac{\partial}{\partial\alpha})},
\end{equation}
\begin{equation}
({g_{1i}})^{+}_{21}=\frac{[iN^{(1/2)}_i-{n}_i/2-(\vec \sigma\times{\bf N}^{(0)})_i]}{2(\alpha-i/2)}{\exp{(\frac{i}{2}\frac{\partial}{\partial\alpha})}},
\end{equation}
and
\begin{equation}
\label{3.20}
({g_{1i}})_{12}^{-}=\frac{\alpha-i}{2\alpha}[-i{n}_i+({\bf n}\times{\vec \sigma})_i]\exp(-\frac{i}{2}\frac{\partial}{\partial\alpha}),
\end{equation}
\begin{equation}
\label{3.21}
({g_{1i}})_{21}^{-}=\frac{[-i{n}_i-({\bf n}\times{\vec \sigma})_i]}{2}{\exp{(-\frac{i}{2}\frac{\partial}{\partial\alpha})}}.
\end{equation}
The  supersymmetry generators  which are inroduced in this section give a connection between the  states for the massive particles. The mass  in the explicit form  appear in the Eqs., (\ref{3.9}), (\ref{3.12}) and (\ref{3.16}) in the operator product ${\exp(-\frac{i}{2}\frac{\partial}{\partial\alpha})}{\exp(\frac{i}{2}\frac{\partial}{\partial\alpha})}$.
 For the mass-zero particles we must exclude this term.
\section{Massless case} 
In order to construct the  supersymmetry generators  for the mass-zero particles we separate   $\widetilde{B}$ and $\widetilde{K}$ in two parts corresponding to the operators ${\exp(\frac{i}{2}\frac{\partial}{\partial\alpha})}$ and ${\exp(-\frac{i}{2}\frac{\partial}{\partial\alpha})}$, respectively
\begin{equation}
\label{4.1}
\widetilde{B}^{+}=\sqrt{{\mu}}D^{\dag(1/2)}({\bf n})\left[1-\frac{i{\vec \sigma}\cdot{\bf L}}{(\alpha-i/2)})\right]\exp(\frac{i}{2}\frac{\partial}{\partial\alpha}),\quad\widetilde{B}^{-}=\sqrt{{\mu}}D^{\dag(1/2)}({\bf n})\exp(-\frac{i}{2}\frac{\partial}{\partial\alpha}),
\end{equation}
\begin{equation}
\label{4.2}
\widetilde{K}^{+}=\sqrt{{\mu}}\left[\frac{\alpha+\imath+i{\vec \sigma}\cdot{\bf L}}{\alpha}\right]D^{(1/2)}({\bf n})\exp(\frac{i}{2}\frac{\partial}{\partial\alpha}),\quad\widetilde{K}^{-}=\sqrt{{\mu}}\left[\frac{\alpha-\imath}{\alpha}\right]D^{(1/2)}({\bf n})\exp(-\frac{i}{2}\frac{\partial}{\partial\alpha}).
\end{equation}
With help of these operators one can  construct two types of representations of the supersymmetric generators: representations with   $\widetilde{B}^{+},\,\widetilde{K}^{+}$ and representations with   $\widetilde{B}^{-},\,\widetilde{K}^{-}$. In  (\ref{4.1})  ) and (\ref{4.2} ) we have introduced a constant ${\mu}$  with dimension of mass in order to deal with dimension operators.

Let us first start with the case of  $\widetilde{B}^{+},\,\widetilde{K}^{+}$.  In  $Q_r$ we replace $\widetilde{B}$ by $\widetilde{B}^{+}$ and $\widetilde{K}$ by $\widetilde{K}^{+}$. The results can be obtained from formulas  (\ref{3.9}) to (\ref{3.18} ) by substituting 
\begin{equation}
\label{4.3}
Q_r{\rightarrow}Q^{+}_r,\quad {Q_{ri}}{\rightarrow}Q^{+}_{ri}=\lbrack{Q^{+}_{r},J_{i}}\rbrack,\quad {G_{ri}}{\rightarrow}G^{+}_{ri}=\lbrack{Q^{+}_{r},N_{i}}\rbrack,
\end{equation}
\begin{displaymath}
{Q^{+}_{1i}}=\left( \begin{array}
{ccc}0 &{\sigma}_i\widetilde{K}^{+}/2\\-\widetilde{B}^{+}{{\sigma}_i}/2 & 0 
\end{array} \right),\quad{G^{+}_{1i}}=\left( \begin{array}
{ccc}0 &\sqrt{{\mu}}[({g_{1i}})^{+}_{12}]D^{(1/2)}({\bf n})\\{\sqrt{\mu}}D^{\dag(1/2)}({\bf n})[({g_{1i}})^{+}_{21}]& 0  
\end{array} \right).
\end{displaymath}
From 
\begin{displaymath}
{Q^{+}_r}^2=2\left( \begin{array}
{ccc}H^{+(0)}_0 & 0\\ 0 &\widetilde{H}^{+(1/2)}_0
\end{array} \right):=2H^{+}_0,
\end{displaymath} 
and
\begin{displaymath}
\{Q^{+}_{r},G^{+}_{ri}\}=-2\imath\left( \begin{array}
{ccc}P^{+(0)}_{0i} & 0\\ 0 & \widetilde{P}^{+(1/2)}_{0i}
\end{array} \right):=-2i{P^{+}_{0i}},
\end{displaymath} 
we obtain the operators (the explicit form of ${{P}^{+(0)}_{0i}}$ and   ${\widetilde{P}^{+(1/2)}_{0i}}$ are given in the Appendix)
\begin{equation}
\label{4.4}
H^{+(0)}_0=\mu[\frac{\alpha(\alpha+\imath)+{\bf L}^2}{2{\alpha}^2}]\exp(i\frac{\partial}{{\partial}{\alpha}}),
\end{equation}
\begin{eqnarray}
\label{4.5}
\widetilde{H}^{+(1/2)}_0=\mu[\frac{\alpha(\alpha+\imath)+(\widetilde{\bf J^{(\frac{1}{2})}})^2}{2({\alpha}^2+\frac{1}{4})}]\exp(i\frac{\partial}{{\partial}{\alpha}}),
\end{eqnarray}
which satisfy the conditions 
\begin{equation}
\label{4.6}
(H^{+(0)}_0)^{2}-({\bf P}^{+(0)}_0)^{2}=0,\quad(\widetilde{H}^{+(1/2)}_0)^{2}-(\widetilde{\bf P}^{+(1/2)}_0)^{2}=0,
\end{equation}
and the commutation relations of the Poincar\'e algebra. Instead of the Eqs., (\ref{3.12}) and (\ref{3.16})  we have
\begin{equation}
\label{4.7}
\{Q^{+}_{ri},Q^{+}_{rk}\}=-H^{+}_0{\delta}_{ik},\quad\{G^{+}_{ri},G^{+}_{rk}\}=-H^{+}_0{\delta}_{ik},
\end{equation}
and one can consider $H^{+(s)}_0$, ${\bf P}^{+(s)}_{0}$  as  the Hamilton  and momentum operators for the mass-zero particles.

In order to go over to the case with $\widetilde{K}^-$ and $\widetilde{B}^-$, we must replace in $Q_r,\,Q_{ri},\,G_{ri}$ the generator $\widetilde{B}$ by $\widetilde{B}^{-}$ and the generator $\widetilde{K}$ by $\widetilde{K}^{-}$.
From ${Q^{-}_1}$
and 
\begin{displaymath}
{Q^{-}_{1i}}=\left( \begin{array}
{ccc}0 &{\sigma}_i\widetilde{K}^{-}/2\\-\widetilde{B}^{-}{{\sigma}_i}/2 & 0 
\end{array} \right),\quad{G^{-}_{1i}}=\left( \begin{array}
{ccc}0 &\sqrt{{\mu}} [({g_{1i}})^{-}_{12}]D^{(1/2)}({\bf n})\\{\sqrt{\mu}}D^{\dag(1/2)}({\bf n})[({g_{1i}})^{-}_{21}]& 0  
\end{array} \right).
\end{displaymath}
we obtain
\begin{equation}
\label{4.8}
{Q^{-}_1}^2=2H^{-}_0,\quad\{Q^{-}_{1},G^{-}_{1i}\}=-2i{P^{-}_{0i}},
\end{equation}
with      
\begin{equation}
\label{4.9}
H^{-(0)}_0=\mu[\frac{(\alpha-\imath)}{2{\alpha}}]
\exp(-i\frac{\partial}{{\partial}{\alpha}}),\quad\widetilde{H}^{-(1/2)}_0=
\mu[\frac{(\alpha-\frac{3i}{2})}{2({\alpha}-\frac{i}{2})}]\exp(-i
\frac{\partial}{{\partial}{\alpha}}),
\end{equation}
\begin{equation}
\label{4.10}
{\bf P}^{-(0)}_{0}={\bf n}H^{-(0)}_0,\quad\widetilde{\bf P}^{-(1/2)}_{0}={\bf n}\widetilde{H}^{-(1/2)}_0.
\end{equation}
For these operators we also have the conditions
\begin{equation}
\label{4.11}
(H^{-(0)}_0)^{2}-({\bf P}^{-(0)}_0)^{2}=0,\quad(\widetilde{H}^{-(1/2)}_0)^{2}-(\widetilde{\bf P}^{-(1/2)}_0)^{2}=0,
\end{equation}
and the commutation relations of the Poincar\'e algebra. Additionally,
\begin{equation}
\label{4.12}
\{Q^{-}_{ri},Q^{-}_{rk}\}=-H^{-}_0{\delta}_{ik},\quad\{G^{-}_{ri},G^{-}_{rk}\}=-H^{-}_0{\delta}_{ik},
\end{equation}
\begin{equation}
\label{4.13}
\lbrack{H^{-}_0,G^{-}_{ri}}\rbrack=0,\quad\lbrack{{\bf P}^{-}_0,Q^{-}_{ri}}\rbrack=0,\quad\lbrack{{\bf P}^{-}_0,G^{-}_{ri}}\rbrack=0.
\end{equation}
For the eigenfunctions of $H^{-(0)}_0$ and ${\bf P}^{-(0)}_{0}$ we may choose $(-\infty<\gamma<\infty)$
\begin{equation}
\label{4.14}
{\Psi}^{-(0)}(\alpha,{\bf n},\gamma,{\bf n}^{'})=\frac{1}{\sqrt{\pi}\alpha}\exp(-\gamma+i\alpha\gamma)\delta({\bf n}-{\bf n}^{'}).
\end{equation}
 Here the  eigenvalues  of $H^{-(0)}_0$ are determined by ${k_0}={\mu}\frac{e^{\gamma}}{2},$ and the eigenvalues of ${\bf P}^{-(0)}_{0}$ by ${\bf k}=k_0{\bf n}^{'}$. For the eigenfunctions of $\widetilde{H}^{-(1/2)}_0$ and $\widetilde{\bf P}^{-(1/2)}_{0}$ we have
\begin{equation}
\label{4.15}
{\Psi}^{-(1/2)}(\alpha,{\bf n},\gamma,{\bf n}^{'})=\widetilde{B}^{-}\frac{{\Psi}^{-(0)}(\alpha,{\bf n},\gamma)}{\sqrt{2k_0}}.
\end{equation}
With help of  $Q^{-}_{r,\,}Q^{-}_{ri}$ and $G^{-}_{ri}$, one can find other  eigenfunctions of $H^{-}_0,\,{\bf P}^{-}_0$.

If we with  $\widetilde{B}^{+}+\widetilde{B}^{-}$ and $\widetilde{K}^{+}+\widetilde{K}^{-}$ return to the massive particles  one can find that in this case  the mass in the Eqs., (\ref{3.9}), (\ref{3.12}) and (\ref{3.16})  may be expressed through the  constant ${\mu}$ for the massless particles. Particularly, for the particles with spin zero 
\begin{equation}
\label{4.16}
  H^{(0)}=H^{+(0)}_0+H^{-(0)}_0,  
\end{equation}
and for the mass 
\begin{equation}
\label{4.17}
{m}=(\widetilde{K}^{+}\widetilde{B}^{-}+\widetilde{K}^{-}\widetilde{B}^{+})/2=\mu.  
\end{equation}
\section{CONCLUSION}
We have shown that in  the generalized Schr\"odinger picture a  relativistic superalgebra may be constructed by using the principal series  of the  unitary representation of the Lorentz group. In the construction  the Poincar\'e algebra for the massive particles with spin zero and spin-1/2  in terms of the invariant parameter ${\alpha}$  and the vector on the  light-cone ${\bf n}$ was used . In this representation we found the explicit form of the supersymmetry generators. For the massless case  we have used  two types of representations of the supersymmetry generators to construct new Hamilton and momentum operators for such particles with spin zero and spin-1/2.
\subsubsection*{ACKNOWLEDGEMENTS}
This work was supported by the Deutsche Forschungsgemeinschaft (No. FR 1560/1-1).
\begin{appendix}
\section*{momentum operators}
The  momentum operators for the massless case  in (\ref{4.6}) may be written as follows
\begin{equation}
{\bf P}^{+(0)}_0={\bf n}{H}^{+(0)}_0-\mu\frac{1}{\alpha}\exp(i\frac{\partial}{{\partial}{\alpha}}){\bf N}^{(0)},
\end{equation}
\begin{eqnarray}
\widetilde{\bf P}^{+(1/2)}_{0}&=&{\bf n}[\widetilde{H}^{+(1/2)}_{0}-\mu\exp(i\frac{\partial}{{\partial}{\alpha}})]\nonumber\\&&-\mu{D^{\dag(1/2)}({\bf n})}\frac{2\alpha({\bf n}{\times}{\bf L})+(\alpha-i/2){\bf n}{\times}{\vec \sigma}+({\bf n}{\vec \sigma}){\bf L}}{2(\alpha^2+1/4)}{D^{(1/2)}({\bf n})}\exp(i\frac{\partial}{{\partial}{\alpha}})].
\end{eqnarray}
\end{appendix}

\end{document}